# Proximity Effect Induced Electronic Properties of Epitaxial Graphene on $Bi_2Te_2Se$

Paengro Lee,[†] Kyung-Hwan Jin,[†] Si Jin Sung,[†] Jin Gul Kim,[†] Min-Tae Ryu,[†] Hee-Min Park,[†] Seung-Hoon Jhi,[†] Namdong Kim,[‡,*] Yongsam Kim,[‡] Seong Uk Yu,[§] Kwang S. Kim,[⊥] Do Young Noh,[∥] and Jinwook Chung[†,*]

[†]Department of Physics, [‡]Pohang Accelerator Laboratory, [§]Department of Chemistry, Pohang University of Science and Technology, Pohang 790-784, Korea, [⊥]Department of Chemistry, Ulsan National Institute of Science and Technology (UNIST), Ulsan 689-798, Korea, and [∥]Department of Physics and Photon Science, Gwangju Institute of Science and Technology, Gwangju 500-712, Korea

*To whom correspondence should be addressed to J.C. (e-mail: jwc@postech.ac.kr) or N.K. (email: east@postech.ac.kr).

**ABSTRACT** We report that the π-electrons of graphene can be spin-polarized to create a phase with a significant spin-orbit gap at the Dirac point (DP) using a graphene-interfaced topological insulator hybrid material. We have grown epitaxial $Bi_2Te_2Se$ (BTS) films on a chemical vapor deposition (CVD) graphene. We observe two linear surface bands both from the CVD graphene notably flattened and BTS coexisting with their DPs separated by 0.53 eV in the photoemission data measured with synchrotron photons. We further demonstrate that the separation between the two DPs, $\varDelta_{D-D}$, can be artificially fine-tuned by adjusting the amount of Cs atoms adsorbed on the graphene to a value as small as $\varDelta_{D-D} = 0.12$ eV to find any proximity effect induced by the DPs. Our density functional theory calculation shows the opening of a spin-orbit gap of ~20 meV in the π-band enhanced by three orders of magnitude from that of a pristine graphene, and a concomitant phase transition from a semi-metallic to a quantum spin Hall phase when $\varDelta_{D-D} \leq 0.20$ eV. We thus present a practical means of spin-polarizing the π-band of graphene, which can be pivotal to advance the graphene-based spintronics.

Topological insulators (TIs)[1–4] and graphene[5–7], despite their different dimensionality and structure, share a common feature revealing the linear Dirac bands with a unique chirality associated with the asymmetry of electron spin and of charge distribution, respectively. Various heterostructures have been studied to create or improve functionality of graphene, for example, a better surface flatness with an improved carrier mobility as reported in the heterostructure of graphene on $h$BN,[8,9] and also for the interface of graphene-TI.[10] Adding a spin degree of freedom to the π-electrons of graphene, in particular, has been a challenge to explore any spin-driven feature that can be exploited in future spintronics.[11–24] The heterostructural interfacing of graphene with TI may bring about new exotic properties mainly stemming from the hybridization of their respective Dirac fermions and the inversion symmetry breaking at the interface to enhance the spin-orbit coupling (SOC) in graphene.[11–13] As is well known, the SOC in graphene through the mixing of π- and σ- electrons is extremely weak,[14–19] and there have been several efforts to enhance the SOC in graphene, for example, by using some adatoms such as thallium, indium, and atomic hydrogen.[20–24]

Here we have devised a new technical means to form a graphene-TI interface, and to enhance the SOC in graphene by bringing the two Dirac points (DPs) together, i.e., the DP-induced proximity effect, without sacrificing the intrinsic nature of graphene.

We have grown a three-dimensional (3D) TI film of $Bi_2Te_2Se$ (BTS) onto a chemical vapor deposition (CVD) graphene, which becomes remarkably flattened upon the growth of BTS with its epitaxial domains increased enough to produce the π-band in our photoemission data. We observe both the topological surface state (TSS) of BTS and the linear π-band of graphene coexisting at the center (Γ-point) of the surface Brillouin zone using angle-resolved photoemission spectroscopy (ARPES) with synchrotron photons. We demonstrate that the two DPs of the π-band of graphene and of the TSS of BTS can be fine-tuned by adjusting the adsorption of external adatoms to enhance the SOC as predicted earlier.[11,12] Our density functional theory (DFT) calculation supports our experimental observations, and further predicts the opening of a significant spin-orbit gap in graphene, when the two DPs overlap at the same symmetry point in the surface Brillouin zone.

**RESULTS AND DISCUSSION**

As schematically drawn in Figure 1a, we have grown an epitaxial BTS thin film on a CVD graphene placed on $SiO_2$ substrate by thermal evaporation with a BTS flake as in our previous work.[10] The BTS film thus grown on graphene and the underlying graphene itself appear to become highly crystalline as seen by the presence of their respective Dirac bands in our ARPES data. Our x-ray data (out-of-plane and in-plane scan) also reveal a well-aligned rotational symmetry between BTS and the underlying SiC substrate in our x-ray data (see Supporting Information). In order to prepare an epitaxial graphene-on-BTS (G/BTS) sample to investigate any proximity effect at the interface, we first formed a thick Au/Ti film layers on top of the BTS. The $SiO_2$ substrate was then mechanically removed from the as-grown sample by using epoxy pasted on top of the Au/Ti film.

A quintuple layer (QL) of BTS consists of five layers of Te-Bi-Se-Bi-Te along the c-axis with lattice constants a = 4.28 Å, and c = 29.84 Å.[25] In Figure 1b, we present top-view of the atomic structure of our G/BTS sample. One finds that the Te layer just below the graphene (a = 2.466 Å) is well aligned with a √3×√3R30° symmetry producing a lattice mismatch 0.2 %, thus suggesting the possibility of a detectable proximity effect with a much enhanced SOC in graphene. Raman spectra shown in Figure 1c obtained with a laser of wavelength 532 nm indicate that graphene covers the whole surface of the BTS. The two peaks at the low frequency region are characteristic Raman peaks of BTS, while G and 2D of graphene are found at the high frequency region as reported earlier.[10] In Figure 1d, we compare C 1s core level of a graphene transferred on $SiO_2$ to that of the G/BTS by fitting with an asymmetric Lorenzian function. The fit curves are shown as black and yellow solid curves for G/BTS and G/$SiO_2$, respectively. The full-width at half-maximum of the G/BTS is found to be 0.402 eV much sharper than 0.847 eV of the graphene/$SiO_2$, indicating the much improved crystallinity of the CVD graphene of the G/BTS with an increased size of epitaxial domains upon growing BTS. For reference, the best linewidth of graphite is 0.36 eV.[26] Moreover, the reduced height distribution found in atomic force microscopy (AFM) images in Figure 1e implies that the graphene sheet of the G/BTS may also have improved electronic properties as reported in other heterostructures using *h*BN.[8]

We have measured the valence band from our G/BTS sample presented in Figure 2a, where one clearly finds the strong π-band and σ-band of graphene as well as relatively weak TSS of the BTS at Γ-point. We also observe the linear Dirac bands at K-point.[27,28] Even though we obtained the valence band by using a photon beam of energy $hv$ = 34 eV with a beam size of < 100 μm, it is rather unusual to observe such well-defined Dirac bands from a CVD graphene. In fact, we failed to observe the π-band with the same ARPES setting from a CVD graphene placed on $SiO_2$ due mainly to the wrinkles and ripples produced during the transfer process.[29] From this observation, we confirm that the CVD graphene becomes significantly flattened in atomic scale upon growing BTS with improved epitaxial domains because of the commensurate lattice matching between graphene and the top Te atomic layer.

As seen in Figure 2b, the graphene π-band of the G/BTS appears to be *p*-doped initially, where the DP locates about 0.2 eV above Fermi energy $E_F$. In order to downshift the DP of graphene below $E_F$, we *n*-doped the graphene by adsorbing Cs on our G/BTS sample (Cs/G/BTS). We find that the DP shifts gradually with increasing Cs dose. With a Cs coverage of $\theta_{Cs}$=0.62 ML, we find the DP at 0.41 eV below $E_F$ (see arrow in Figure 2c). The amount of charge transfer calculated from the shift of DP by using the relation $n=(E_D-E_F)^2/[\pi(\hbar v_F)^2]$, we obtain n~$2.6\times10^{13}$ $cm^{-2}$ with $E_D-E_F$=0.61 eV and $v_F$~$1\times10^6$ m/s.[30,31] Interestingly, from the magnified and contrast-enhanced bands near $E_F$ shown in Figure 2d (and also reproduced in Figure 2f and 2g), we observe both the weak π-band of graphene and the TSS of BTS coexisting at Γ. The Λ-shaped band at Γ is apparently a zone-folded one of the π-band at the K-point due to the underlying √3×√3R30° symmetry. Notice also the two Fermi surface (FS) contours in Figure 2d of the graphene π-band (inner blue circle) and of the TSS of BTS (outer red circle).

From the contrast-enhanced band images at Γ from the G/BTS (Figure 2f), one clearly observes the relatively weak zone-folded π-band of graphene and the TSS of BTS. Figure 2g shows the same band image from the Cs/G/BTS, which reveals significant changes in the band structure especially for the position of the DPs of both surface bands. Note that we have overlapped our DFT calculated bands, blue for graphene and red curves for BTS, discussed below for comparison. In order to determine the position of the DPs, we have fitted intensity profiles of the two momentum distribution curves from each sample with two Gaussian functions. The best-fit TSS ($\pi$-) band marked by the yellow (white) dashed lines shows how its DP and Fermi velocity are changed upon the adsorption of Cs on the G/BTS. The DP of the $\pi$-band ($D_G$) is downshifted by 0.65 eV from $D_G$ = +0.18 eV above $E_F$ while that of BTS ($D_{TI}$) by 0.24 eV from $D_{TI}$ = -0.35 eV below $E_F$ at 0.62 ML of Cs to make their separation, $\varDelta_{D-D}$ = 0.12 eV for the Cs/G/BTS, much reduced from $\varDelta_{D-D}$ = 0.53 eV of the G/BTS. With the reduced $\varDelta_{D-D}$, we notice that the Fermi velocity of graphene is also reduced slightly from $0.85 \times 10^6$ m/s of the G/BTS to $0.78 \times 10^6$ m/s of the Cs/G/BTS though theory predicts much more.[11] Since the DP is found to shift nearly in proportion to the Cs coverage by a different amount for each surface band, one can delicately control the Cs coverage to locate the two DPs at the same energy near $\Gamma$. As discussed below, our theory predicts that such a proximity of the two DPs may significantly enhance the SOC to open a giant spin-orbit gap in graphene.

The Cs-induced changes in the core levels (C 1$s$, Se 3$d$, Te 4$d$, and Bi 5$d$) of the Cs/G/BTS presented in Figure 3 indicate that the different amounts of charge transfer from Cs to graphene and TI is the source of driving the proximity of the DPs.[32] These core levels were taken with a synchrotron photon beam of energy $hv$ = 510 eV, and the spectra other than C 1$s$ are not shown in Figure 3. The C 1$s$ core level taken from the Cs/G/BTS (Figure 3a) at $\theta_{Cs}$=0.62 ML appears quite asymmetric in shape,[26,32,33] and has three components, G from $sp^2$−bonded carbons, S1 from $sp^3$−bonded carbons at the edges or from defective dangling bonds, and S2 from the Cs-adsorbed graphene.[26,32] The origin of S2, located 1.02 eV from G, derived from the carbon atoms in the Cs-C bondings, is not associated with any chemical shift due to the adsorption of Cs since no such a shift is found in Cs $4d$ or in any other surface core levels. We thus ascribe the S2 to the intrinsic electron-hole excitations as in K-[26] or Cs-adsorbed graphene[33] due to the rearrangement of electrons near Fermi level in the presence of holes created by the ejected electrons. As observed for the Cs-induced changes of the DPs in Figure 2, one also notices similar shifts in binding energy induced by Cs, $\Delta E_b$= +0.62 eV for the G-component of C 1$s$, and $\Delta E_b$=+0.20 eV for Bi 5$d$ at $\theta_{Cs}$=0.62 ML, but almost none for other core levels. This provides a strong clue for the bonding configuration of Cs on graphene. Our DFT calculation shows that the TSS of the BTS is localized mostly at outer QLs of the BTS and the charge density ρ at the surface comprises electrons mainly from Bi and Te. Moreover, the maximum binding energy for Cs is found from the hollow (H) site, rather than the bridge (B) or on-top (T) site. We thus conclude that Cs adatoms occupy the H-site just above the Bi atoms. This explains why the DP of TSS is so sensitive to the adsorption of Cs as seen in Figure 2 through a charge mixing dominantly not only with C but also with Bi. Such a charge mixing is quite apparent in Figure 3c, where the charge difference $\Delta\rho=\rho_{(Cs/G/BTS)} - \rho_{(BTS)} - \rho_{(G)} - \rho_{(Cs)}$ induced by the Cs adsorption is depicted as a function of the depth (z) along the surface normal. Regions in red (+) and blue (-) depict the isosurfaces with $\Delta\rho=\pm 0.005$ electrons/Å$^3$, respectively. One notices that the charge modulation induced by Cs is mostly localized at the interfacial region around graphene with the transferred electrons distributed in between Cs and the graphene layer. Our calculation also reveals that some of core electrons at the interface are depleted to occupy the graphene π-band to produce the core level shift toward the shallow side rather than the tighter side as the screening effect describes.[32,34]

Unfortunately, the valence band of the Cs/G/BTS in Figure 2g becomes much diffused because of the positional disorder among the randomly distributed Cs adatoms bonded at the H-sites. In order to find any DP-driven proximity effect blurred in our band data, we have carried out DFT band calculations by adopting the same method described earlier.[11] We find quite interesting features especially on graphene suggesting a proximity effect when the two DPs of the surface states both from BTS and graphene come close together within 0.2 eV.[11,12]

Figure 4a shows a progressive change of the two DPs of the Cs/G/BTS as a function of charge transfer from Cs ($\Delta e$) per 100 carbon atoms. The two DPs are found to approach each other with increasing $\Delta e$, and opens a spin-orbit gap ($\Delta_{SO}$) when $\Delta_{D-D} \leq 0.20$ eV (inset). The charge transfer $\Delta e$ is adjusted by varying the atomic distance ($d_{Cs}$) between Cs and graphene as shown in Figure 4b. Figure 4c predicts that at the critical coverage of Cs or equivalently when $\Delta_{D-D} \leq 0.20$ eV, a spin-orbit gap of 20 meV opens as a result of the enhanced SOC in the π-electrons of graphene, and at the same time the semi-metallic (SM) phase of graphene changes into a quantum spin Hall (QSH) phase when the SOC exceeds the staggered on-site potential stemming from the asymmetry of the sublattices.[11] This result indicates that graphene, a truly 2D material, not only exhibits fascinating electronic properties by itself but also, when hybridized with TI materials, can be utilized as a platform to explore the QSH effects.

**CONCLUSION**

In summary, we have prepared a graphene-TI heterostructure (G/BTS) by epitaxially growing BTS on graphene, and investigated exotic electronic properties of the graphene unique only at the interface by using ARPES. We observe that two DPs approach closely each other as Cs atoms are adsorbed on graphene. We thus demonstrate that the $\Delta_{D-D}$ can be delicately tuned by adjusting the amount of Cs adatoms on graphene. Our DFT calculation predicts interesting features in graphene of this heterostructure such as the opening of a giant spin-orbit gap (20 meV) in the $\pi$-band, and a phase transition from a semi-metallic to a quantum spin Hall phase when $\Delta_{D-D} \leq 0.20$ eV. Though these theoretical predictions are not clearly discernable in our band images, mainly due to the diffused band images upon dosing Cs on graphene, we present a prototypical graphene-TI heterostructure as a candidate to be efficiently utilized in future graphene-based spin-sensitive nanodevices.

**EXPERIMENTAL SECTION**

**ARPES measurements** We have carried out our ARPES measurements at the 4A1 BL of Pohang Light Source (PLS) for the valence band with synchrotron photons of $h\nu$ = 34 eV. The incoming photon beam size was < 100 μm and photoelectrons were detected by using a Scienta SES-2002 electron energy analyzer under the base pressure of ~ $5\times10^{-11}$ Torr. Our core-level data were obtained at the 8A2 BL of PLS with photons of $h\nu$ = 510 eV using a Scienta SES-100 analyzer. We have used a SAES getter source to deposit Cs atoms on graphene while maintaining the substrate at 30 K. The evaporation rate was repeatedly monitored with a quartz microbalance. As a reference, we calibrated a monolayer coverage of Cs as the coverage when the Bi 4*f* core-level peak is reduced down to half to that from the clean BTS, considering the escape depth of 1 nm for photoelectrons, which is about the thickness of the first QL of the BTS.

**DFT calculation**   In order to understand and quantify our measured bands of the G/BTS material, especially when the two DPs are close enough, we have performed DFT band calculations in the plane-wave basis within the generalized gradient approximation. A cutoff energy of 400 eV was used for the expansion of wave functions. The k-point meshes of 11×11×1 were used for the sampling of the Brillouin zone. The G/BTS surface structure was simulated in a supercell with graphene on one surface of the slab of five QLs of the BTS, and a vacuum layer of 20 Å-thick between the cells to minimize artificial inter-cell interactions. We used the experimental lattice constant (a = 4.28 Å) of the BTS[25] and graphene was strained by 0.2 % to match the substrate.

**Supporting Information.**   Atomic force microscopy (AFM) data of G/SiO$_2$, G/BTS, and of graphene and x-ray data from G/BTS are presented in Fig. S1 and Fig. S2, to show the crystallinity of the G/BTS and the more flattened domains of carbons in G/BTS than in G/SiO$_2$, respectively. Brief description of fitting method with fit-results of our measured ARPES data is given in Fig. S3 to identify the surface bands. Additional band calculations with and without SOC are provided in Fig. S4 to support the gap originating from the SOC rather than sublattice asymmetry. Additional band data showing the spin-induced gap as a function of SOC strength are shown in Fig. S5 to support the presence of the QSH state of graphene. The supporting material is available free of charge *via* the Internet at http://pubs.acs.org.

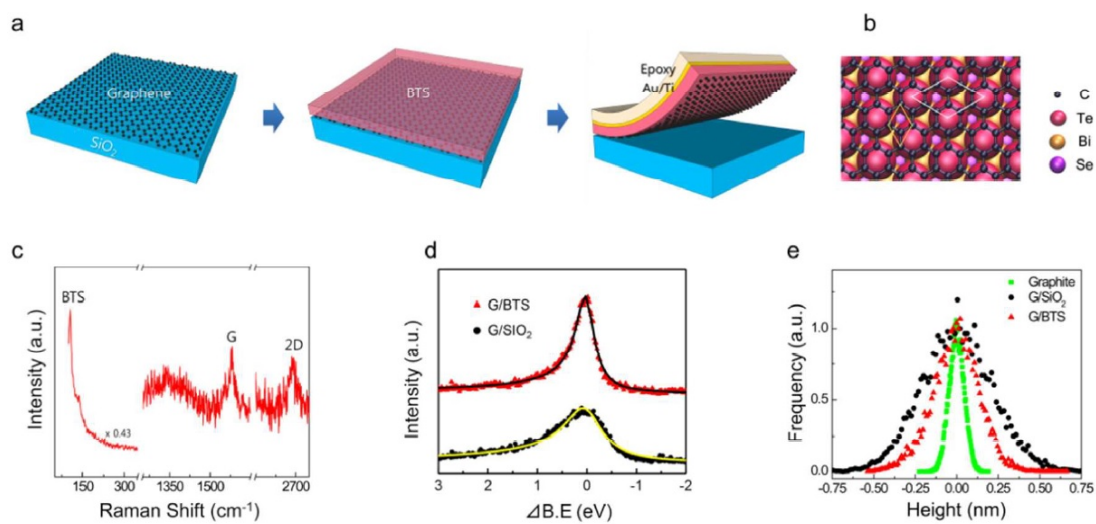

**Figure 1.** Epitaxial graphene on BTS. (a) A crystalline film of BTS was first grown on a CVD graphene placed on a SiO$_2$ substrate, and a thick Au/Ti film layers were then formed on top of the BTS. The SiO$_2$ substrate was mechanically removed from the as-grown sample by using epoxy pasted on top of the Au/Ti film. (b) Top view of the atomic structure of the G/BTS. Note the √3×√3R30° symmetry between graphene and top Te atoms. (c) Raman spectra from the G/BTS sample showing the characteristic peaks of BTS (105, 139 cm$^{-1}$) together with G and 2D (1582, 2689 cm$^{-1}$) peaks of graphene. (d) Comparison of C 1s core levels of G/BTS and G/SiO$_2$ revealing the nearly doubled size of epitaxial domains of the G/BTS. The black (red) curve is the fit-curve of the data for G/BTS (G/SiO$_2$). (e) Height distributions of graphene measured by AFM showing improved flatness in G/BTS compared to that in G/SiO$_2$.

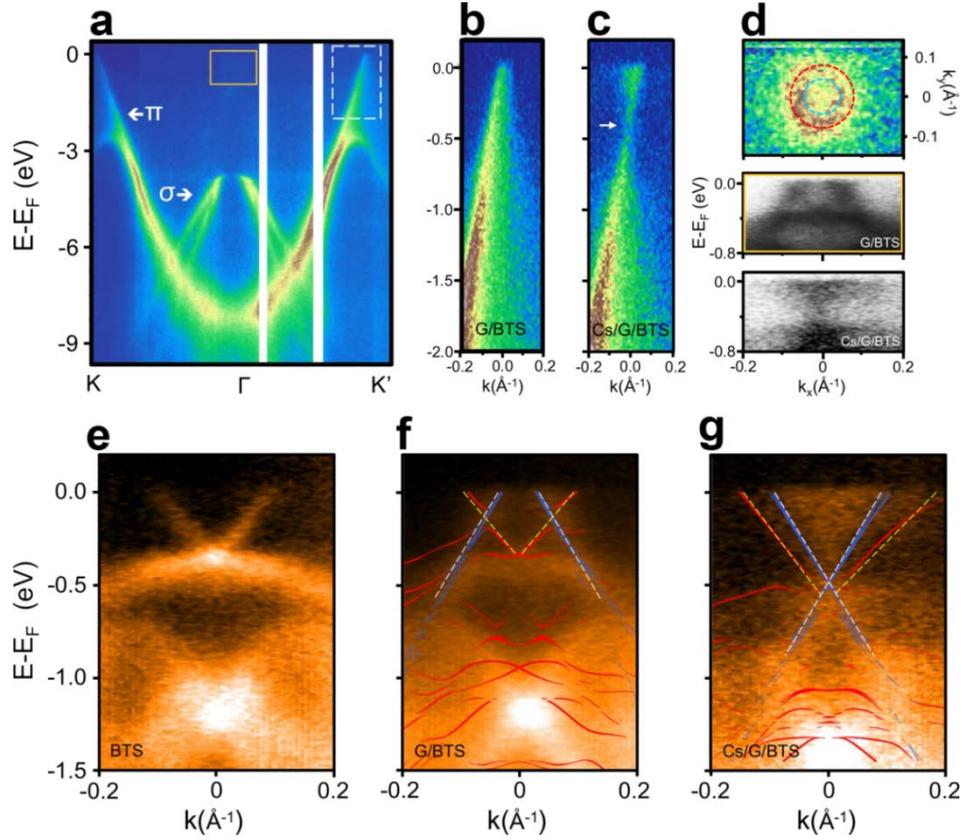

**Figure 2.** Changes in valence band of the G/BTS with the Cs dose obtained with a photon energy of hv = 34 eV at temperature 24 K. (a) Valence band of the clean G/BTS where the magnified images of the dashed box are shown in (b) and (c) from clean G/BTS and Cs adsorbed G/BTS (Cs/G/BTS). One notices a *p*-doped π-band of graphene in (b), which is shifted down by *n*-doping from Cs adsorbates to move the DP 0.4 eV below $E_F$ as shown in (c). (d) Fermi surface contours (upper) from the G/BTS. The inner (blue) circle is from the graphene π-band, while outer (red) circle from the TSS of BTS extracted from the contrast-enhanced band image (lower) of the yellow box in (a). The contrast-enhanced valence band near the TSS of the BTS is shown in (e), that of the G/BTS in (f), and that of the Cs/G/BTS in (g). In (f) and (g), DFT calculated bands are overlapped, where the measured (calculated) TSS of BTS and the π-band of graphene are denoted by yellow (red) and white (blue) dotted (solid) curves, respectively.

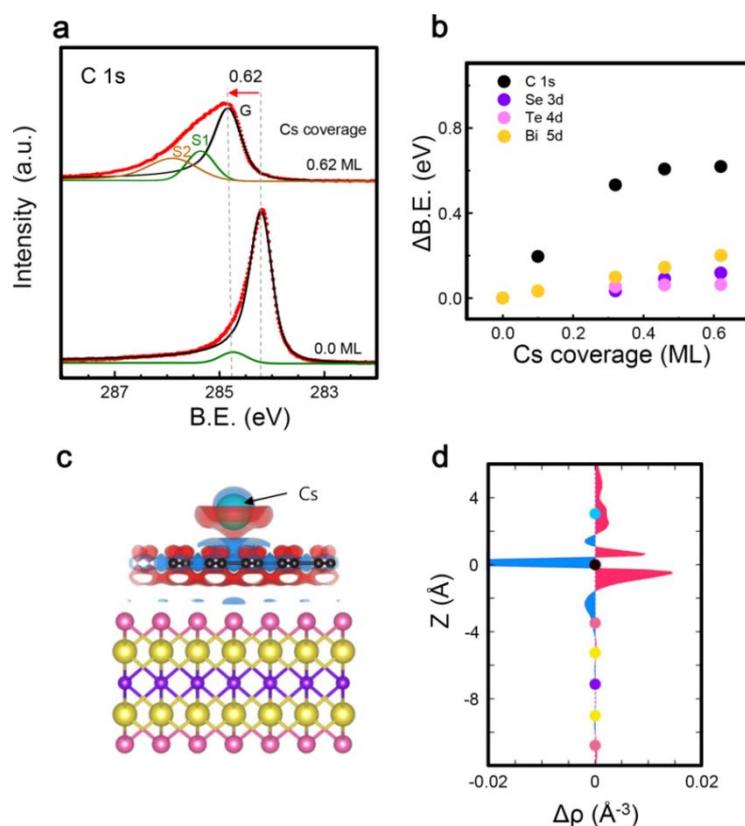

**Figure 3.** Bonding configuration of Cs adatom on graphene. (a) Change in the C 1s with Cs adsorption. All core level spectra were obtained with photon energy of hv =510 eV. The main component G of C 1s shifts by 0.62 eV upon Cs dose ($\theta_{Cs}$=0.62 ML) accompanied by two additional components, S1 from defective carbons and S2 from Cs-C bonding. (b) Changes in the binding energy of other core levels (Se 3d, Te 4d, and Bi 5d) with increasing $\theta_{Cs}$. (c) Side view of the charge density induced by one Cs adatom in a 2×2 G/BTS supercell. (d) The charge difference $\Delta\rho(z)$ averaged along the surface normal with graphene located at z=0. Red (blue) areas correspond to the accumulation (depletion) of electrons.

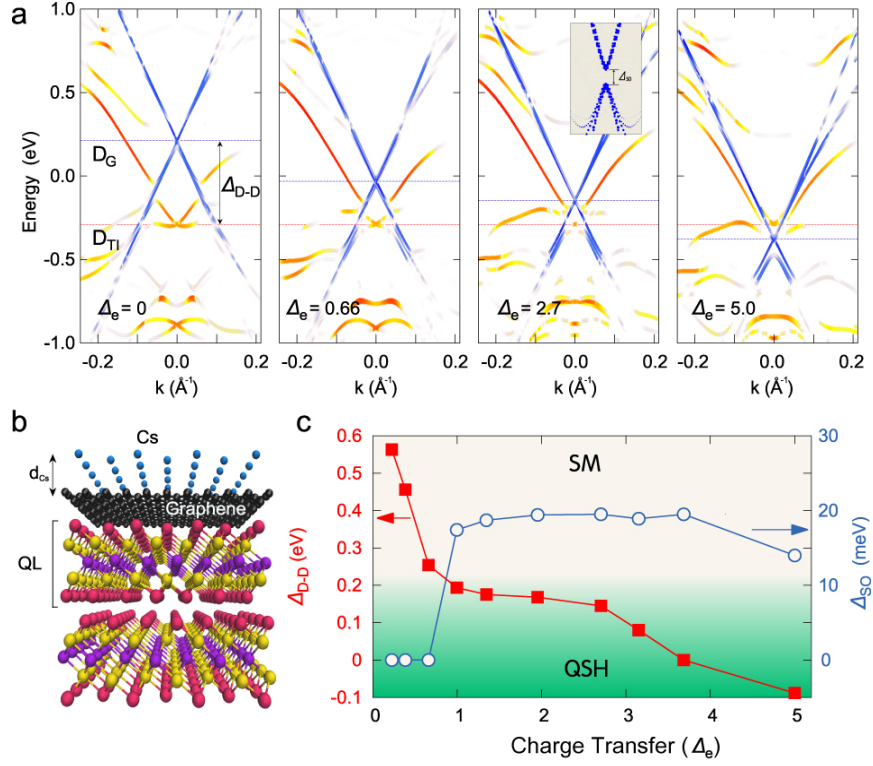

**Figure 4.** Progressive changes of the G/BTS with increasing Cs coverage estimated by the amount of transferred electrons ($\Delta e$) per 100 carbon atoms. (a) Calculated bands of the G/BTS as a function of $\Delta e$. States originating from graphene (BTS) is colored in blue (red), where the DP for graphene (BTS) is denoted by $D_G$ ($D_{TI}$). (b) Atomic structure of the Cs-adsorbed G/BTS. The amount of charge transfer, $\Delta e$, from Cs is controlled by varying the Cs bonding distance ($d_{CS}$). (c) The energy separation between the two Dirac cones ($\Delta_{D-D} = D_G - D_{TI}$) and the intrinsic spin-orbit gap ($\Delta_{SO}$) of graphene as a function of $\Delta e$. One finds that the energy separation gradually decreases with increasing Cs and a giant spin-orbit gap of graphene of about 20 meV opens when $\Delta_{D-D} \leq 0.2$ eV, where the semi-metallic (SM) phase of graphene ($\Delta_{D-D} > 0.2$ eV) turns into a quantum spin Hall (QSH) phase ($\Delta_{D-D} < 0.2$ eV).